\documentclass[pre,twocolumn,showpacs]{revtex4}
\usepackage{epsfig}

\begin{document}

\title{Water waves over a time-dependent bottom: 
Exact description for 2D potential flows}
\author{V.~P. Ruban}
\email{ruban@itp.ac.ru}
\affiliation{Landau Institute for Theoretical Physics,
2 Kosygin Street, 119334 Moscow, Russia} 

\date{\today}

\begin{abstract}
Two-dimensional potential flows of an ideal fluid  with a free surface 
are considered in situations when shape of the bottom depends on time
due to external reasons. 
Exact nonlinear equations describing surface waves 
in terms of the so called conformal variables are derived for an arbitrary 
time-evolving bottom parameterized by an analytical function. 
An efficient numerical method for the obtained equations is suggested.
\end{abstract}

\maketitle


\section{Introduction}

The theory of water waves now is among the most actively developing 
branches of the hydrodynamic science (see, for instance, 
\cite{Z1999,A-R,Kirby,CH1999, BT1996,ONS2000,TS1996,
Ovs,DKSZ96,DLZ95,ZD96,Lvov,DZK96,D2001,ZDV2002,R2004PRE},
and many references therein).
In particular, two-dimensional (2D) flows of an ideal incompressible fluid
with a free surface are very attractive for theoretical
study, since they possess many essential features in common with real water 
waves, and simultaneously are relatively simple for analytical and numerical 
treatment. A powerful mathematical support for the 2D theory is provided by 
analytical functions and the corresponding conformal mappings.
The original idea of using conformal mappings was intended
for waves on the deep water or over a static horizontal bottom: a geometrically 
simple region occupied  by resting fluid 
(it is the lower half-plane \cite{Ovs,DKSZ96,DLZ95,ZD96,Lvov,D2001,ZDV2002}, 
or a horizontal stripe \cite{DZK96}) 
is mapped onto the flow region with a disturbed time-dependent free 
surface, so that the real axis is transformed into the moving boundary. 
With such conformal ``straightening'' of the free boundary, 
it is possible to obtain an exact analytical description for  surface waves.
The corresponding (1+1)-dimensional nonlinear equations of motion are very 
convenient for numerical simulations by spectral methods.
Important achievements in this direction were reported in works 
\cite{DKSZ96,DLZ95,ZD96,DZK96,Lvov,D2001,ZDV2002}, including analytical and
numerical results. A generalization has 
been recently made to the case when the bottom shape 
is still static but strongly inhomogeneous in space \cite{R2004PRE}.
 
The purpose of present paper is to derive exact equations of motion
for the free boundary in 2D ideal flows,  valid with arbitrary non-stationary 
bottom profile (Sec.II), and to suggest a numerical method for solving the 
obtained highly nonlinear equations (Sec.III). This problem, besides being 
interesting by itself, is also important 
for studying such phenomena as tsunami generation by earthquakes and other 
similar processes in natural and laboratory conditions.

\section{Exact nonlinear equations}

Let us start our consideration with basic definitions.
We will describe a nonstationary flow region ${\cal D}_z$ in 
$(x,y)$-plane (with $y$-axis up-directed) by an analytical function 
$x+iy=z(w,t)$, where $t$ is the time, and the complex 
variable $w=u+iv$ occupies in $(u,v)$-plane the fixed infinite horizontal
stripe ${\cal D}_w:-\infty<u<+\infty, 0<v<1$.
Position of the free surface $\partial{\cal D}^{(s)}_z$  
will be given in parametric form by the formula
\begin{equation}\label{surface_shape}
X^{(s)}(u,t)+iY^{(s)}(u,t)\equiv Z^{(s)}(u,t)=z(u+i,t),
\end{equation}
while the bottom profile $\partial{\cal D}^{(b)}_z$will be determined by
\begin{equation}\label{bottom_shape}
X^{(b)}(u,t)+iY^{(b)}(u,t)\equiv  Z^{(b)}(u,t)=z(u,t).
\end{equation}
Inasmuch as the velocity potential $\varphi(x,y,t)$ satisfies the
Laplace equation $\varphi_{xx}+\varphi_{yy}=0$, in ${\cal D}_z$  
the analytical complex potential
$\tilde\phi(z,t)=\varphi(x,y,t)+i\theta(x,y,t)$ is defined, with 
$\theta(x,y)$ being the harmonically conjugate for $\varphi(x,y)$.  
We deal below with the complex analytical function 
$\phi(w,t)=\tilde\phi(z(w,t),t)$ defined in ${\cal D}_w$ and taking 
the boundary values as written below:
\begin{equation}\label{boundary_potential}
\phi(u+i,t)\equiv \Phi^{(s)}(u,t),\qquad \phi(u,t)\equiv \Phi^{(b)}(u,t).
\end{equation}
The velocity components are determined by the usual relations
\begin{equation}\label{VxVy}
V_x-iV_y=\varphi_x-i\varphi_y=d\tilde \phi/dz =\phi'(w,t)/z'(w,t),
\end{equation}
where $(\dots)'$ denotes derivative over the complex variable $w$. 

Now we proceed to equations of motion.
First of all, it is the dynamic Bernoulli equation on the free surface,
\begin{equation}\label{Bernoulli_Dz}
\mbox{Re}(\tilde\phi_t(z,t)) +
\frac{|d\tilde\phi/dz|^2}{2}+g\,\mbox{Im\,}z =0,
\qquad z\in \partial{\cal D}^{(s)}_z,
\end{equation}
where $g$ is the gravitational acceleration. This equation can be easily 
re-written in terms of the conformal variables as follows:
\begin{equation}\label{Bernoulli_Dw}
\mbox{Re}\left(\Phi^{(s)}_t-\Phi^{(s)}_u\frac{Z^{(s)}_t}{Z^{(s)}_u}\right) +
\frac{|\Phi^{(s)}_u|^2}{2|Z^{(s)}_u|^2}+g\,\mbox{Im\,}Z^{(s)} =0.
\end{equation}

Two other equations of motion are the kinematic conditions on the free surface
and on the bottom, expressing the fact that the boundary motion is  
determined by the normal component of the velocity field.
In the conformal variables these conditions take form
\begin{equation}\label{kinematic_Dw_s}
\mbox{Im}\left(Z^{(s)}_t\bar Z^{(s)}_u\right)=-\mbox{Im\,}\Phi^{(s)}_u,
\end{equation}
\begin{equation}\label{kinematic_Dw_b}
\mbox{Im}\left(Z^{(b)}_t\bar Z^{(b)}_u\right)=-\mbox{Im\,}\Phi^{(b)}_u,
\end{equation}
where $\bar Z$ denotes complex conjugate value.

We should take into account that $\Phi^{(s)}(u,t)$ and $\Phi^{(b)}(u,t)$
are not mutually independent, but they are related to each other by a linear
transform. Indeed, the following formula for analytical continuation is valid:
\begin{equation}\label{analytic_continuation}
\phi(w,t)=\int_{-\infty}^{+\infty} \Phi^{(b)}_k(t)e^{ikw}\frac{dk}{2\pi}.
\end{equation}
Here the Fourier transform $\Phi^{(b)}(u)\mapsto\Phi^{(b)}_k$ 
is defined in the usual way:
\begin{equation}\label{Phi_b_k}
\Phi^{(b)}_k(t)=\int\Phi^{(b)}(u,t)e^{-iku} du.
\end{equation}
Therefore at the free surface, where $w=u+i$, we have
\begin{equation}\label{Phi_s_Phi_k}
\Phi^{(s)}(u,t)=\int e^{-k}\Phi^{(b)}_k(t)e^{iku}\frac{dk}{2\pi}.
\end{equation}
It is a proper point here to introduce some linear operators, necessary for
further exposition. These linear operators
$\hat S$, $\hat R$, and $\hat T=\hat R^{-1}$  
are diagonal in Fourier representation:
\begin{equation}\label{SR}
 S_k={1}/{\cosh(k)},\quad 
 R_k=i\tanh(k),\quad  T_k=-i\coth(k).
\end{equation}
It will be convenient for us to write $\Phi^{(b)}(u,t)$ in the form
\begin{equation}\label{Phi_b}
\Phi^{(b)}(u,t)=\hat S \psi(u,t) -i (1-i\hat R) \hat\partial_u^{-1}f(u,t),
\end{equation}
with $\psi(u,t)$ and $f(u,t)$ being some unknown real functions. 
Then for $\Phi^{(s)}(u,t)$ we will have the formula
\begin{equation}\label{Phi_s}
\Phi^{(s)}(u,t)=(1+i\hat R)\psi(u,t) -i \hat S  \hat\partial_u^{-1}f(u,t).
\end{equation}
Later we use in equations the complex function 
\begin{equation}\label{Psi_def}
\Psi(u,t)\equiv (1+i\hat R)\psi(u,t).
\end{equation}

Now it is necessary to look at the conformal mapping $z(w,t)$
with more attention. Since the bottom motion is assumed to be prescribed, 
this function  has the following structure (compare with \cite{R2004PRE}):
\begin{equation}\label{z_zeta_w}
z(w,t)=Z(\zeta(w,t),t),
\end{equation}
where a known analytical function $Z(\zeta,t)$ determines a conformal mapping
for the upper half-plane of an intermediate complex variable $\zeta$ onto
the infinite region above the bottom in $z$-plane (or, at least, the 
mapping $Z(\zeta,t)$ should have no singularities within a sufficiently 
wide horizontal stripe above the real axis in $\zeta$-plane, so there 
should be enough ``free space'' for large-amplitude waves in $z$-plane).
The intermediate analytical function $\zeta(w,t)$ takes real values at the real
axis. Therefore we may write 
\begin{equation}\label{zeta_def}
\zeta(w,t)=\int \frac{a_k(t)}{\cosh(k)}e^{ikw}\frac{dk}{2\pi}, 
\qquad a_{-k}=\bar a_k
\end{equation}
where $a_k(t)$ is the Fourier transform of a real function $a(u,t)$.
On the bottom $\zeta(u,t)=\hat S a(u,t)$, thus the parametrically given curve
\begin{equation} \label{Zb}
Z^{(b)}(u,t)=Z(\hat S a(u,t),t), \qquad -\infty < u < +\infty,
\end{equation}
coincides with the prescribed curve $X+iY=Z(s,t), -\infty < s < +\infty$
(these two curves differ from each other just by parameterization).
At the free surface 
\begin{equation}\label{xi_def}
\zeta(u+i,t)\equiv\xi(u,t)=(1+i\hat R)a(u,t),
\end{equation}
\begin{equation}\label{Zs} 
Z^{(s)}(u,t)=Z(\xi(u,t),t).
\end{equation}

Since our purpose is to derive equations of motion for the unknown  
functions $\Psi(u,t)$, $f(u,t)$, and $\xi(u,t)$, we have to 
substitute expressions (\ref{Phi_b}), (\ref{Phi_s}), (\ref{Zb}),
(\ref{xi_def}), and (\ref{Zs}) into Eqs.(\ref{Bernoulli_Dw}), 
(\ref{kinematic_Dw_s}), (\ref{kinematic_Dw_b}). 
Eq.(\ref{kinematic_Dw_b}) is simple and does not need any trick.
Eq.(\ref{kinematic_Dw_s}) (divided by $|Z^{(s)}_u|^2$) takes the form
\begin{equation}\label{kinematic_Dw_s_2}
\mbox{Im\,}\left(\frac{\xi_t}{\xi_u}\right)+
\mbox{Im\,}\left(\!\frac{Z_t(\xi,t)}{Z_\xi(\xi,t)\xi_u}\!\right)
=\frac{-\mbox{Im\,}\Psi_u +\hat S f}{|Z_\xi(\xi,t)\xi_u|^2}.
\end{equation}
From here we express $\xi_t$ (see Eq.(\ref{xi_t}) below and compare with
\cite{R2004PRE}) and substitute
proper expressions into Eq.(\ref{Bernoulli_Dw}), in order to find $\psi_t$.
Simple transformations, similar to those in work \cite{R2004PRE}, 
lead to the following equations:
\begin{eqnarray}
f\!\!&=&\mbox{Im\,}
\Big(Z_t(s,t)\bar Z_s(s,t)\hat S a_u\Big)\Big|_{s=\hat S a},\quad
a=\mbox{Re\,}\xi,
\label{f}\\
\xi_t\!\!&=&\!\!-\xi_u\,(\hat T\!+\!i)\!\!\left[
\frac{\mbox{Im\,}\Psi_u -\hat S f}
{|Z_\xi(\xi,t)\xi_u|^2}+\mbox{Im}\left(\!
\frac{Z_t(\xi,t)}{Z_\xi(\xi,t)\xi_u}\!\right)\right]\!\!,
\label{xi_t}\\
\Psi_t\!\!&=&\!\!-\Psi_u\,(\hat T\!+\!i)\!\!\left[
\frac{\mbox{Im\,}\Psi_u -\hat S f}
{|Z_\xi(\xi,t)\xi_u|^2}+\mbox{Im}\left(\!
\frac{Z_t(\xi,t)}{Z_\xi(\xi,t)\xi_u}\!\right)\right]\nonumber\\
&&+(1\!+\!i\hat R)\Bigg[
\mbox{Re}\left(\frac{\Psi_u Z_t(\xi,t)}{Z_\xi(\xi,t)\xi_u}\right)
-\frac{|\Psi_u|^2 - (\hat S f)^2}{2|Z_\xi(\xi,t)\xi_u|^2}\nonumber\\
&&\qquad\qquad\qquad\qquad\qquad
-g\, \mbox{Im\,}Z(\xi,t)\Bigg].
\label{Psi_t}
\end{eqnarray}
This system of equations is the main result of present paper. It provides exact
description for potential water waves over a time-dependent bottom determined by
an analytical function $Z(\zeta,t)$. It should be noted that capillarity 
effects can be easily included here by adding the term 
$$
\sigma |\hat \partial_u Z(\xi(u))|^{-1}
\mbox{Im\,}[(\hat \partial_u)^2Z(\xi(u))/\hat \partial_u Z(\xi(u))]
$$
(proportional to the surface curvature) to the gravity term
$-g\, \mbox{Im\,}Z(\xi)$.

The above equations are valid not only for infinitely extended configurations
(when the fluid covers the whole bottom), 
but also for the flows with one or two non-stationary shore edges. 
However, asymptotic properties of the intermediate 
mapping $\zeta(w,t)$ are different in these cases. So, with two edges 
$\zeta(u,t)\to C_\pm(t)$ as $u\to\pm\infty$, while for edge-less configurations
the corresponding limits are $\zeta(u,t)\to \pm\infty$.

\section{Numerical method}

Equations (\ref{f}), (\ref{xi_t}), (\ref{Psi_t}), 
though look quite complicated, in many cases are convenient for numerical 
simulations, as described below.  
\begin{figure}
\begin{center}
  \epsfig{file=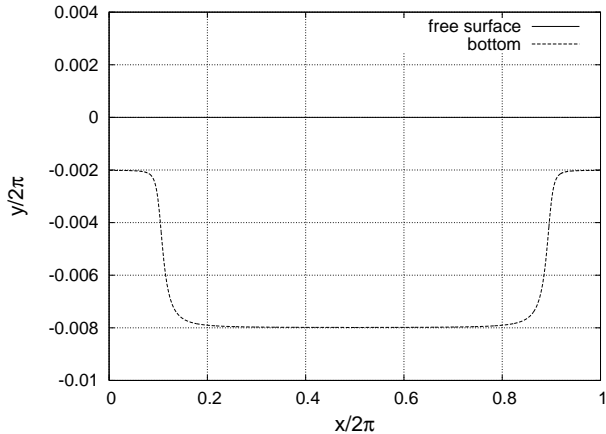,width=84mm}   
\end{center}
\caption{\small (i) Initial bottom profile and horizontal free surface. 
The velocity field is everywhere zero.} 
\label{i-t0}
\end{figure}
\begin{figure}
\begin{center}
  \epsfig{file=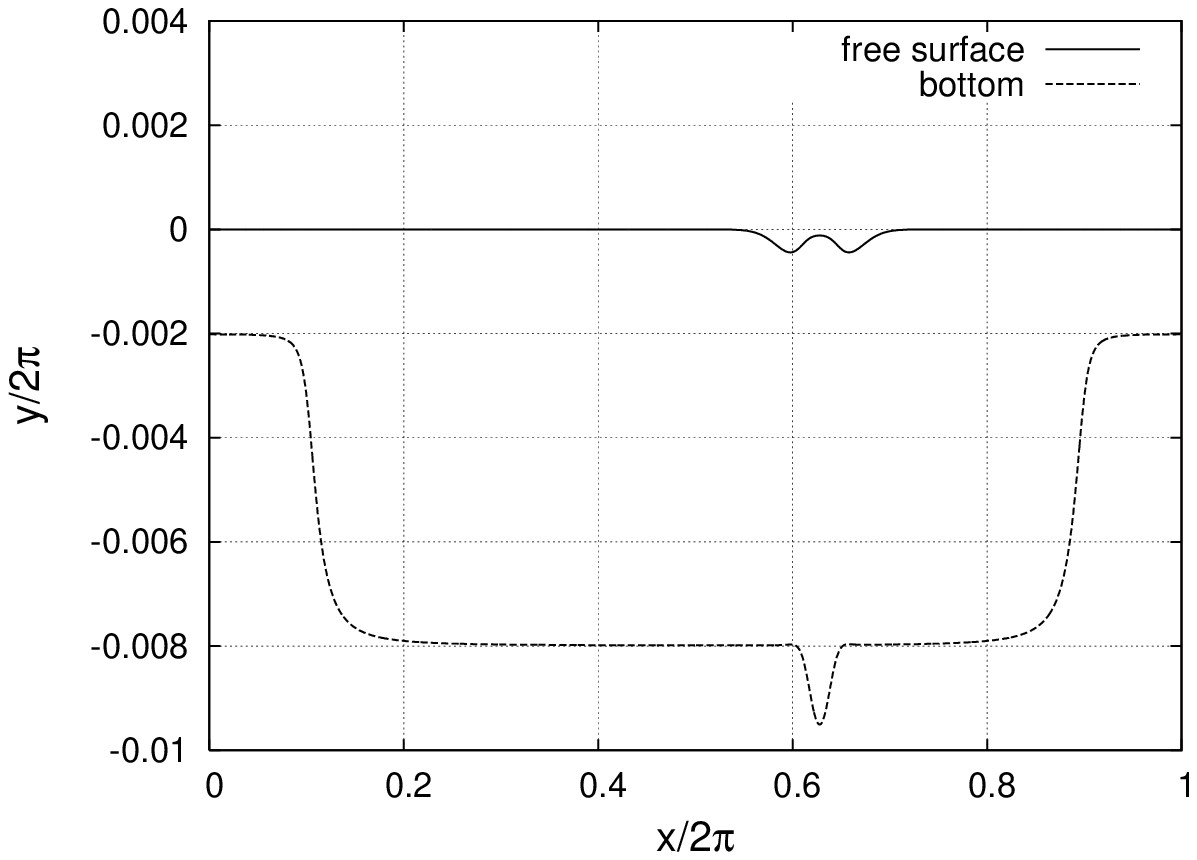,width=84mm}
  \epsfig{file=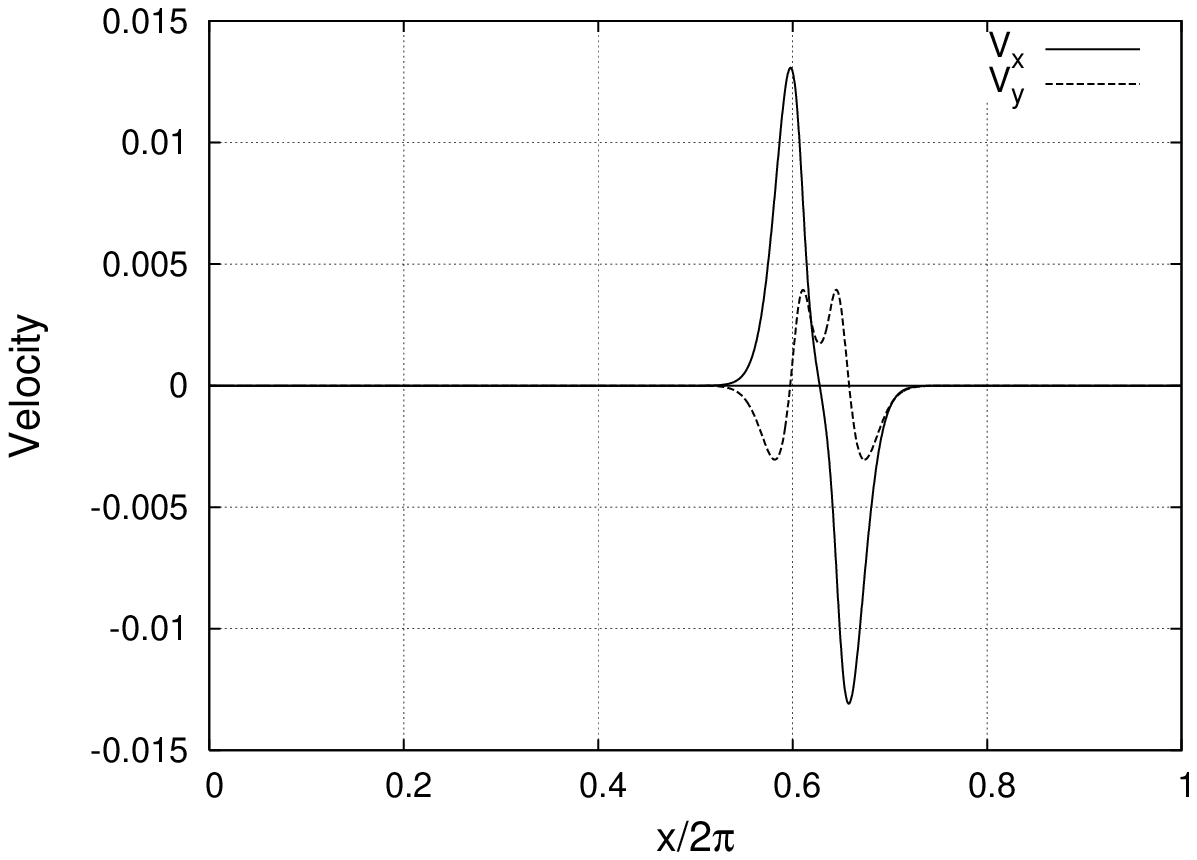,width=84mm}
\end{center}
\caption{\small (i) $t=2$: The cavity on the bottom has been almost formed.} 
\label{i-t1}
\end{figure}
\begin{figure}
\begin{center}
  \epsfig{file=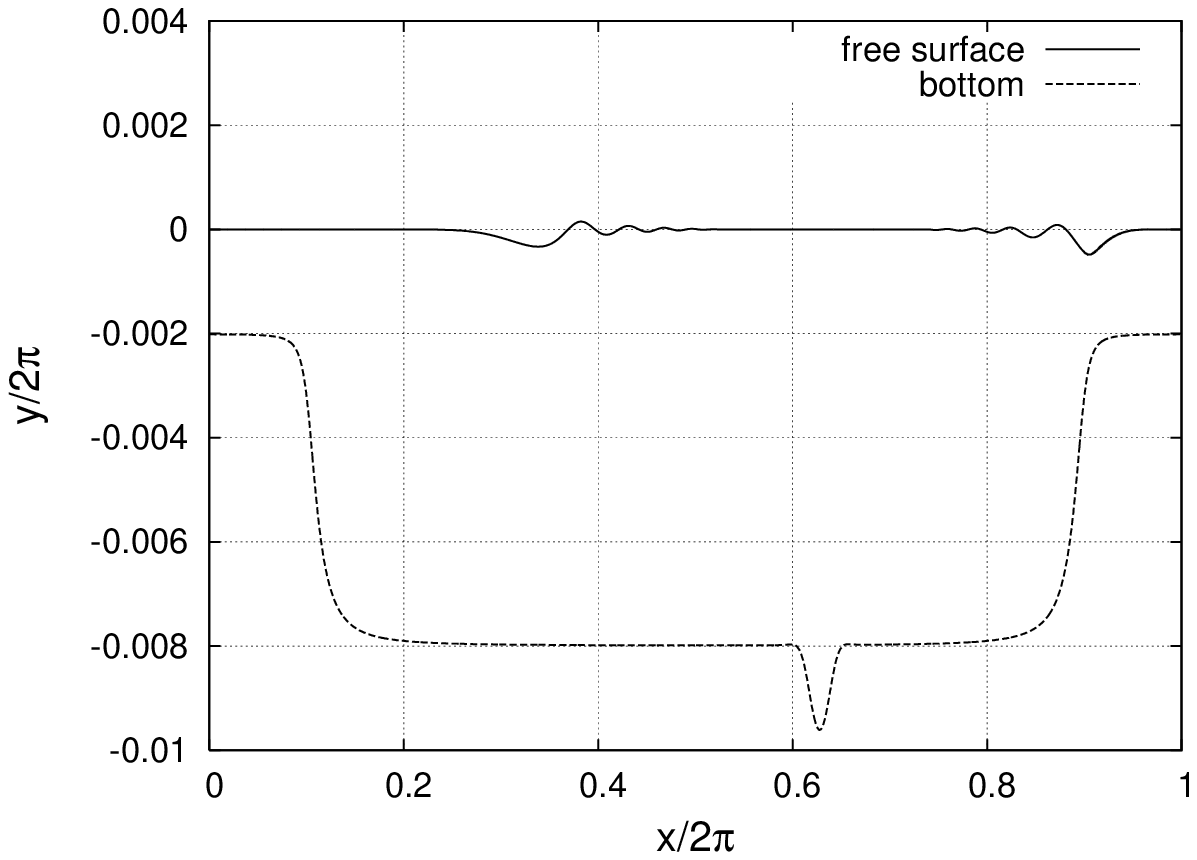,width=84mm}
  \epsfig{file=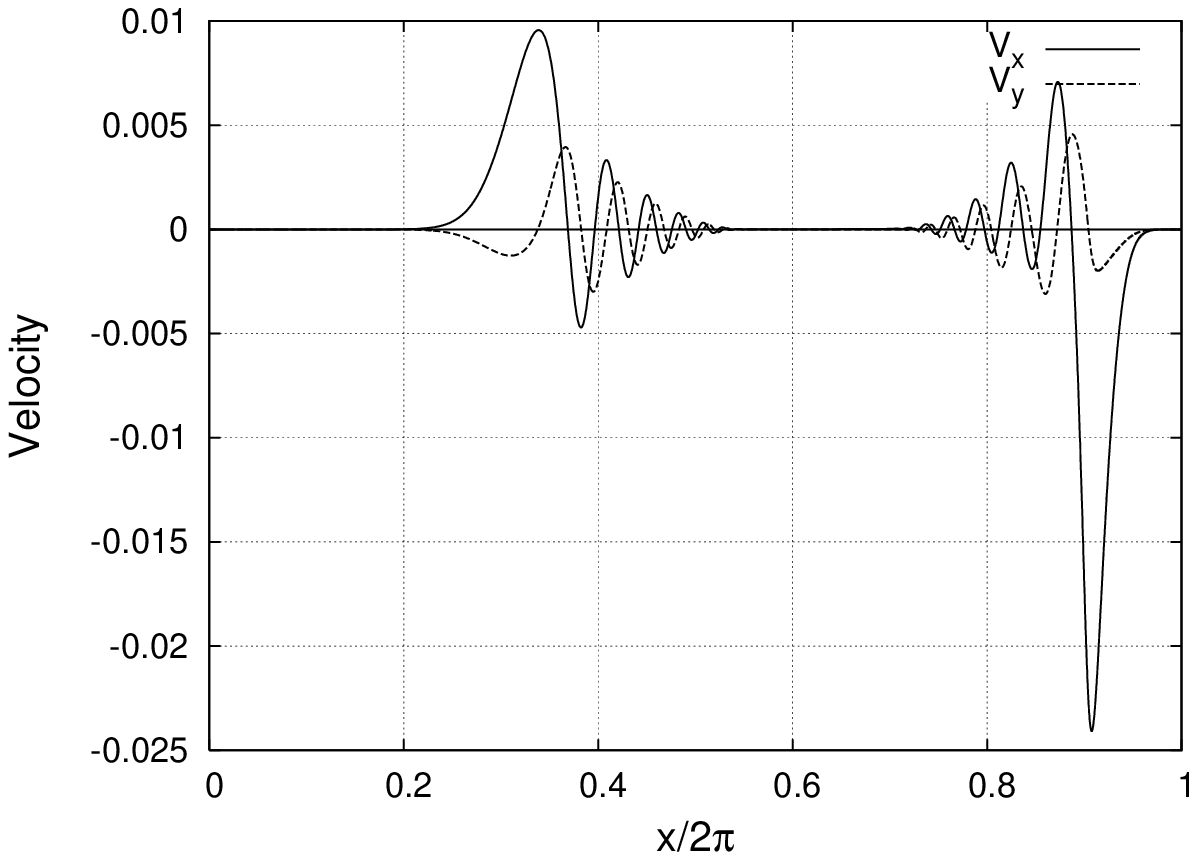,width=84mm}
\end{center}
\caption{\small (i) $t=10$: The right-propagating group of
waves is approaching the shallow region.} 
\label{i-t2}
\end{figure}
\begin{figure}
\begin{center}
  \epsfig{file=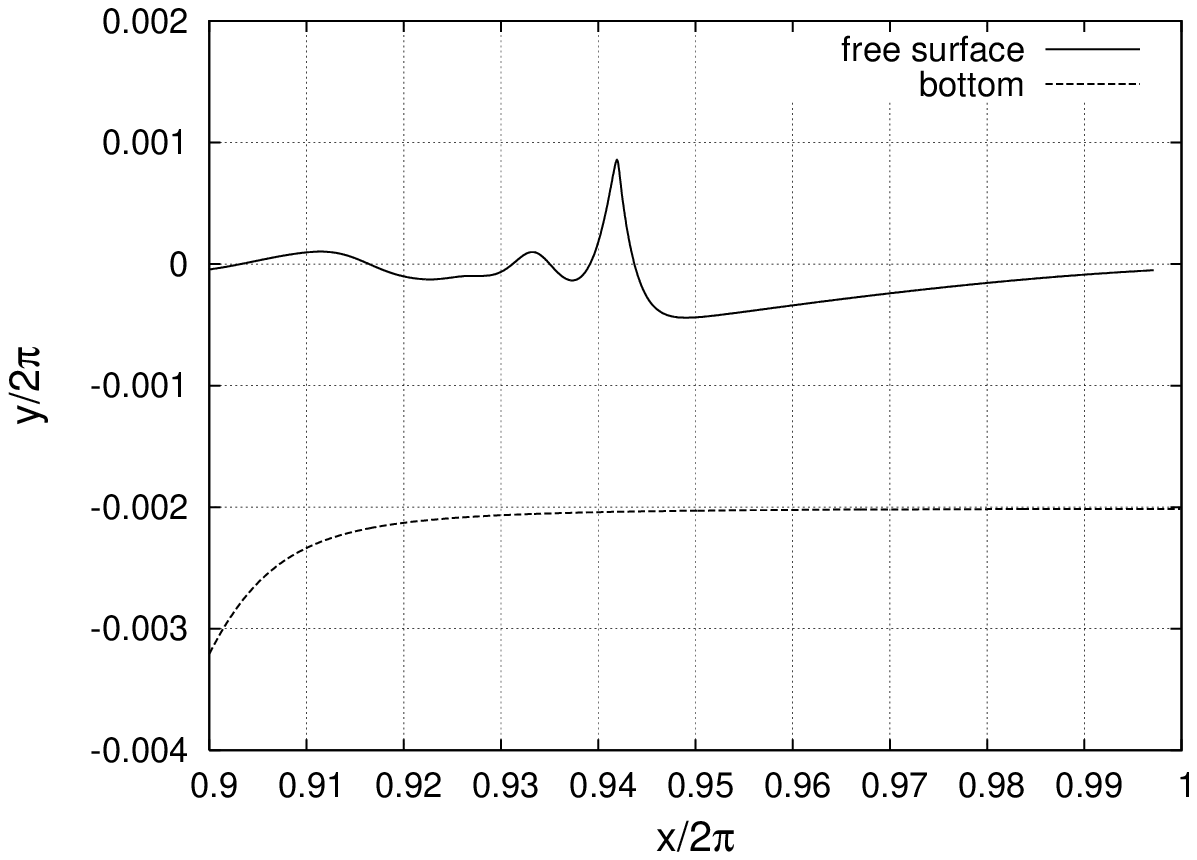,width=84mm}
  \epsfig{file=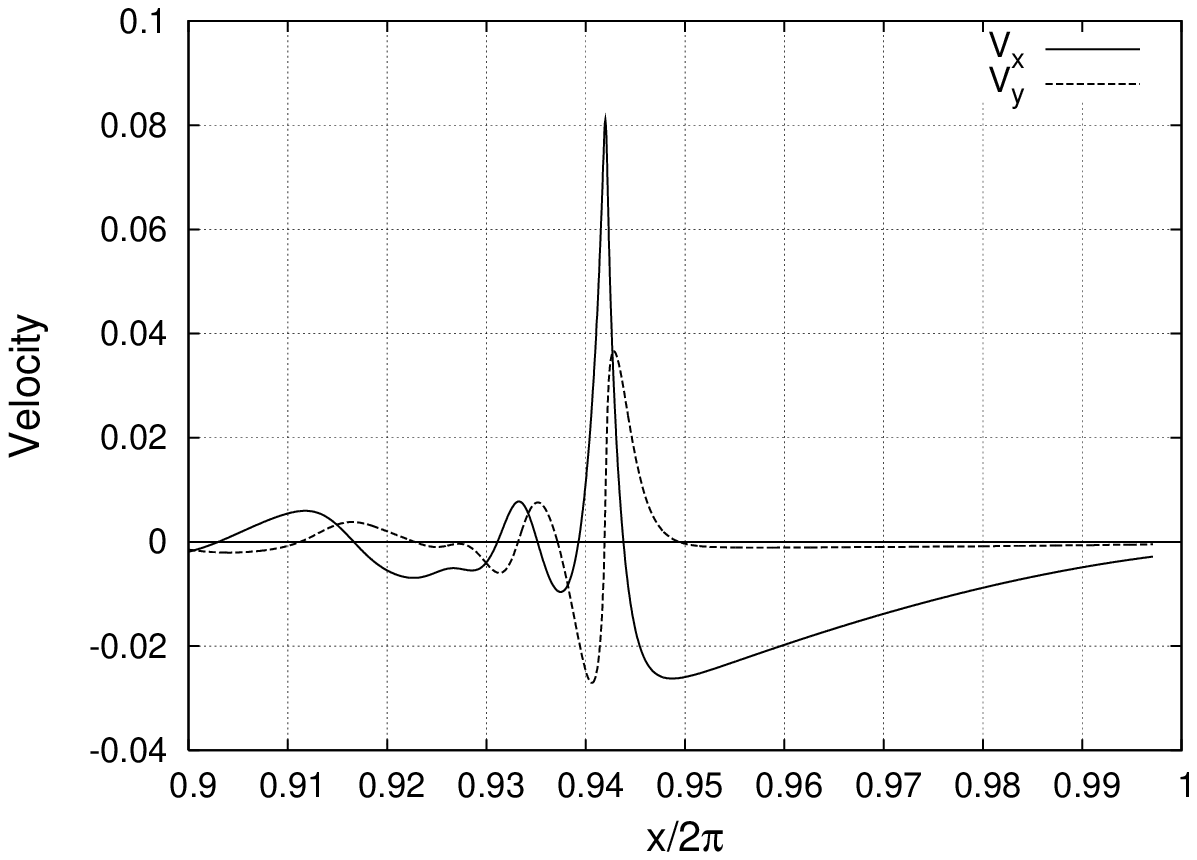,width=84mm}   
\end{center}
\caption{\small (i) Breaking leading wave at $t\approx 13.18$} 
\label{i-t3}
\end{figure}

Instead of the function $\Psi(u,t)$, we deal with another function
(similar to that proposed in \cite{D2001,ZDV2002}), 
\begin{equation}\label{beta_def}
\beta(u,t)=\Psi_u(u,t)/\xi_u(u,t),
\end{equation}
because this choice results in better numerical stability 
(concerning the problem of optimal variables see \cite{LZ2004,D2004}). 

Let us consider a class of space-periodic flows (without shore edges!) 
over $x$-periodic profiles of the bottom. With a proper choice 
for the length and time scales, we may write $g=1$, and 
$Z(\zeta+2\pi,t)=2\pi+Z(\zeta,t)$. As it was pointed in \cite{R2004PRE},
in space-periodic flows the variable $u$ comes into solutions in the 
combination $\vartheta=u \alpha(t)$, where a real function $\alpha(t)$
depends on time because  the operator $\hat T$ is singular
on small wave-numbers. Besides $\alpha(t)$, solutions are determined 
by two other real functions, $\rho(\vartheta,t)$ and $\chi(\vartheta,t)$, 
both having period $2\pi$ on $\vartheta$ variable:
\begin{eqnarray}
\rho(\vartheta,t)=\!\sum_{m=-\infty}^{+\infty}\!\rho_m(t)\exp(im\vartheta),
\quad \rho_{-m}(t)=\bar\rho_m(t),&&\\
\chi(\vartheta,t)=\!\sum_{m=-\infty}^{+\infty}\!\chi_m(t)\exp(im\vartheta),
\quad \chi_{-m}(t)=\bar\chi_m(t),&&
\end{eqnarray}
\begin{eqnarray}
\xi(\vartheta,t)&=&\vartheta+i\alpha(t)
+\sum_{m=-\infty}^{+\infty}\frac{2\rho_m(t)\exp(im\vartheta)}
{1+\exp(2m\alpha(t))}\nonumber\\
&=&\vartheta+i\alpha(t)+(1+i\hat{\mathsf R}_\alpha)\rho(\vartheta,t),  
\label{xi_rho}
\end{eqnarray}
\begin{equation}
\beta(\vartheta,t)=\sum_{m=-\infty}^{+\infty}
\frac{2\chi_m(t)\exp(im\vartheta)}
{1+\exp(2m\alpha(t))}=(1+i\hat{\mathsf R}_\alpha)\chi(\vartheta,t).
\label{beta_chi}
\end{equation}
The linear operator $\hat{\mathsf R}_\alpha$  is diagonal in the discrete
Fourier representation:
${\mathsf R}_\alpha(m)=i\tanh(\alpha m)$.
Let us introduce the following quantities:
\begin{eqnarray}
F\!\!\!&=&\mbox{Im}
\Big[Z_t(s,t)\bar Z_s(s,t)
(1+\hat {\mathsf S}_\alpha \partial_{\vartheta}\rho)\Big]
\Big|_{s=\vartheta+\hat {\mathsf S}_\alpha \rho},
\label{F}\\
U\!\!\!&=&\!\!(\hat{\mathsf T}_\alpha\!+\!i)\!\left[
\frac{\mbox{Im\,}(\beta\xi' )\!-\!\hat {\mathsf S}_\alpha F}
{|Z_\xi(\xi,t)\xi'|^2}+\mbox{Im}\!\left(\!
\frac{Z_t(\xi,t)}{Z_\xi(\xi,t)\xi'}\!\right)\right]\!\!,
\\
W\!\!\!&=&\!\!\frac{|\beta\xi'|^2\!-\!(\hat {\mathsf S}_\alpha F)^2}
{2|Z_\xi(\xi,t)\xi'|^2}+\mbox{Im}Z(\xi,t)
-\!\mbox{Re}\!\left(\!\!\frac{\beta Z_t(\xi,t)}
{Z_\xi(\xi,t)}\!\!\right)\!\!,
\end{eqnarray}
where ${\mathsf S}_\alpha(m)=1/\cosh(\alpha m)$,
$\xi'\equiv\partial_{\vartheta}\xi=
1+(1+i\hat{\mathsf R}_\alpha)\partial_\vartheta\rho$. 
The linear operator $\hat{\mathsf T}_\alpha$ is regular. In the discrete
Fourier representation it is defined as follows:
\begin{eqnarray}
{\mathsf T}_\alpha(m)&=&-i\coth(\alpha m), \qquad m\not=0;\nonumber\\
&=& 0,\qquad\qquad \qquad\quad m=0.
\end{eqnarray}
Equations of motion for the real functions  $\alpha(t)$, $\rho(\vartheta,t)$, 
and $\chi(\vartheta,t)$ follow from (\ref{f}), (\ref{xi_t}), (\ref{Psi_t})
and have the form
\begin{eqnarray}
\dot\alpha(t)&=&-\frac{1}{2\pi}\int_0^{2\pi}\mbox{Im\,}(U)
d\vartheta,
\label{dot_alpha}\\
\dot \rho(\vartheta,t)&=&-\mbox{Re}\left(U\xi'\right),
\label{dot_rho}\\
\dot \chi(\vartheta,t)&=&
-\mbox{Re}\left(U\beta'+\frac{1}{\xi'}(1+i\hat{\mathsf R}_\alpha)
\partial_{\vartheta}W\right),
\label{dot_chi}
\end{eqnarray}
where $\beta'\equiv\partial_{\vartheta}\beta=
(1+i\hat{\mathsf R}_\alpha)\partial_\vartheta\chi$.

The above system (\ref{dot_alpha})-(\ref{dot_chi})
can be efficiently simulated by the spectral method, 
since the multiplications are simple in $\vartheta$-representation, 
while the linear operators $\hat{\mathsf S}_\alpha$,
$\hat{\mathsf R}_\alpha$, and
$\hat{\mathsf T}_\alpha$ (also $\vartheta$-differentiation) 
are simple in Fourier representation.
The integration scheme can be based, for instance, on the Runge-Kutta 
4-th order accuracy method, similarly to works \cite{ZDV2002, R2004PRE}.
Efficient subroutine libraries for the fast Fourier transform are now
available. In the numerical experiments reported here, the FFTW library 
was used  \cite{fftw3}. The employed Fourier harmonics had numbers $m$ 
in the limits $-16000<m<16000$ (the size of arrays for the Fourier transform
was even larger, $N=2^{16}$), thus the obtained solutions are very accurate
(compare with \cite{R2004PRE}). 

\subsection{First numerical experiment}

In the first numerical experiment [referred to as (i)], 
the shape of the bottom was determined by 
analytical function $Z(\zeta,t)={\cal B}^{(i)}(\zeta-i\alpha_0,t)$, where 
$\alpha_0=\pi/100\approx 0.0314$,
and function ${\cal B}^{(i)}(q,t)$ is expressed as follows:
\begin{eqnarray}
{\cal B}^{(i)}(q,t)&=&q
+i\Delta\ln\left(\frac{i\sin q+\sqrt{\epsilon+\cos^2q}}
{\sqrt{1+\epsilon}}\right)\nonumber\\
&+&\frac{iAt^4}{\tau^4+t^4}\exp[-C(1+\cos(q-\delta))].
\end{eqnarray}
Here the parameters are $\Delta=0.6$, $\epsilon=0.01$, $A=-0.008$, $\tau=1.0$,
$C=500.0$, and $\delta=0.16\pi$. The time-independent terms in the 
above formula describe the bottom profile as shown in Fig.\ref{i-t0}, 
with deep and shallow regions.
The time-dependent term in ${\cal B}(q,t)$ corresponds to formation 
of a cavity on the bottom. The initial values were taken
$\rho(\vartheta,0)=0$, $\chi(\vartheta,0)=0$, and $\alpha(0)=\alpha_0$, 
thus giving the horizontal free surface and zero velocity field at $t=0$.
The corresponding numerical results 
are presented in Figs.\ref{i-t0}-\ref{i-t3}
in terms of the dimensionless quantities $z=x+iy$ and 
$V=[\beta\xi'-i\hat{\mathsf S}_\alpha F]/[Z_\xi(\xi,t)\xi']=V_x-iV_y$.
This numerical experiment can be viewed as a rough 
model for a tsunami wave generation, subsequent propagation, and final breaking.
Indeed, the inhomogeneous bottom displacement is an analog of an earthquake
resulting in two opposite-propagating groups of surface waves.
When one of the groups first approaches a shallow region, the crest of its 
leading wave becomes higher and sharper, and finally singular. In real world 
instead true singularity one can observe vortices, splash, and foam on the
crest. Further treatment of this situation is not possible within present ideal 
2D theory, since 3D effects and dissipative processes become important.

\subsection{Second numerical experiment}
\begin{figure}
\begin{center}
  \epsfig{file=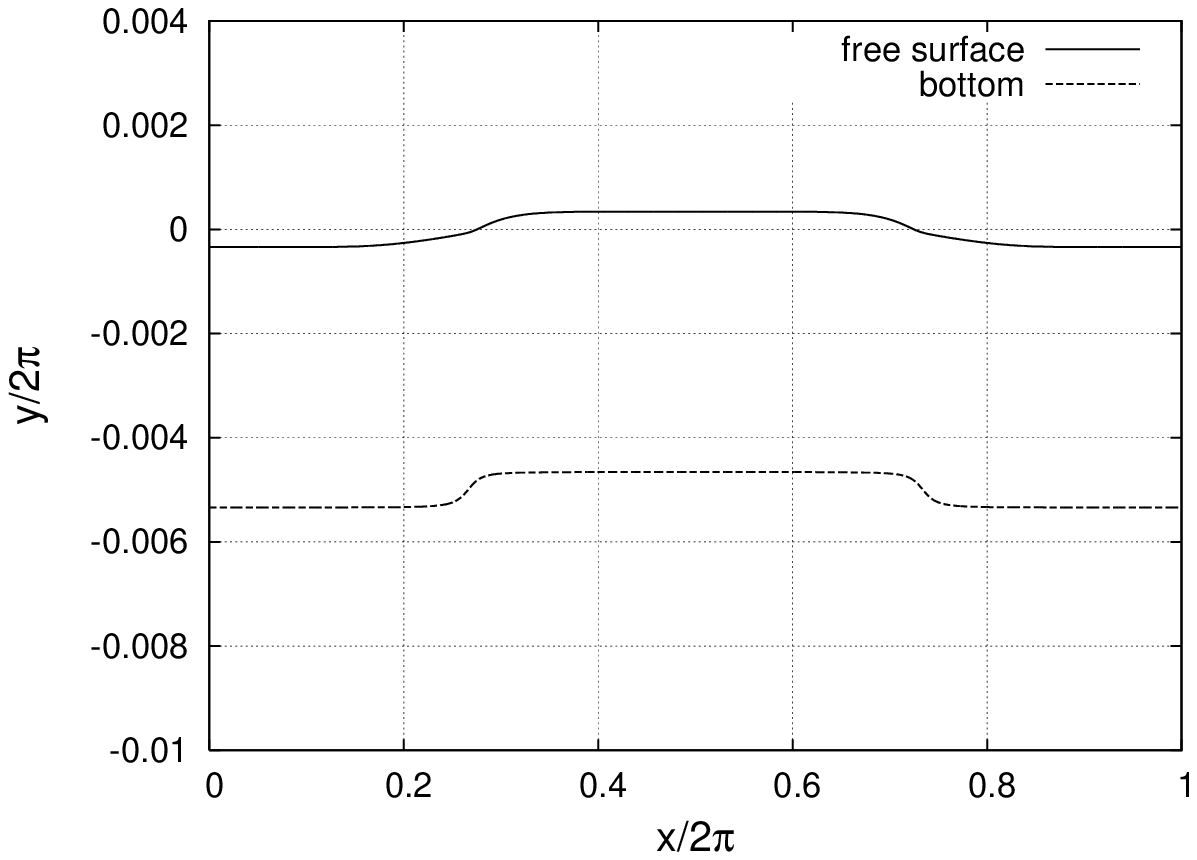,width=74mm}
  \epsfig{file=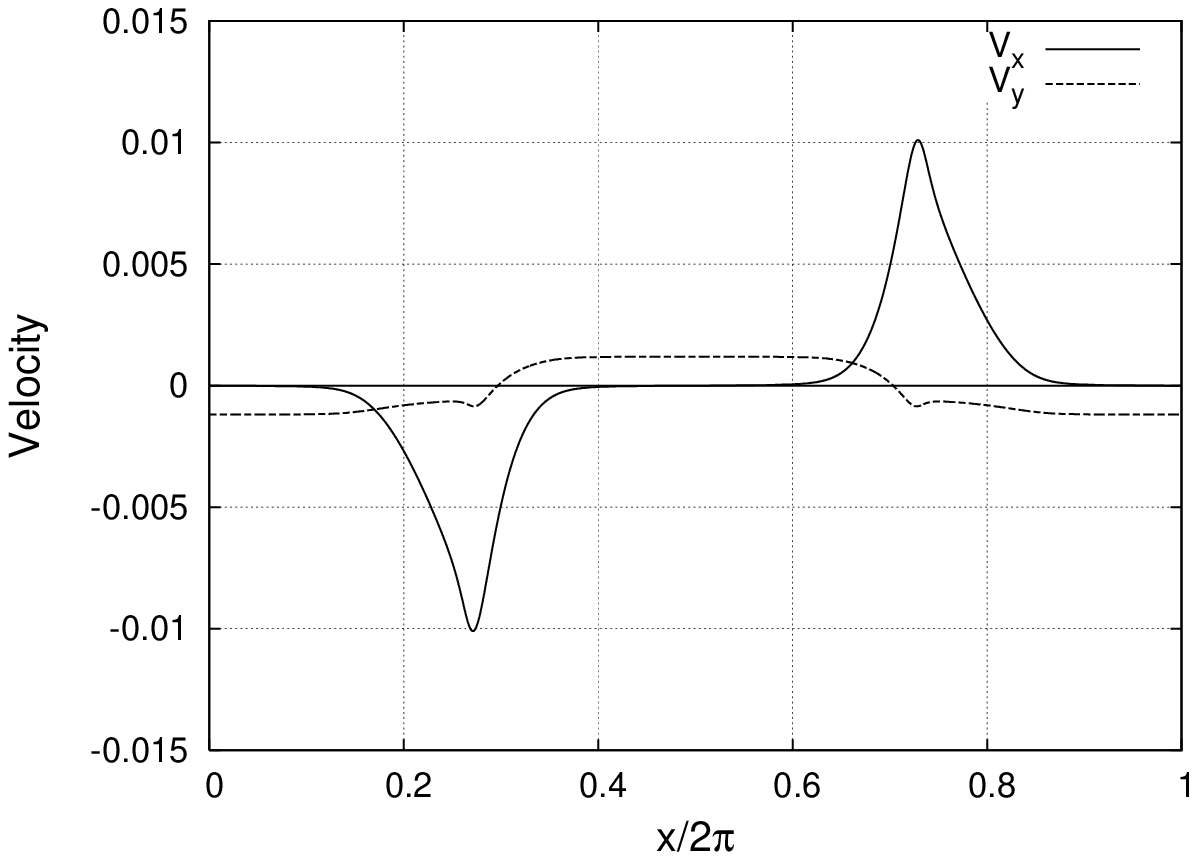,width=74mm}
\end{center}
\caption{\small (ii) $t=5$} 
\label{ii-t1}
\end{figure}
\begin{figure}
\begin{center}
  \epsfig{file=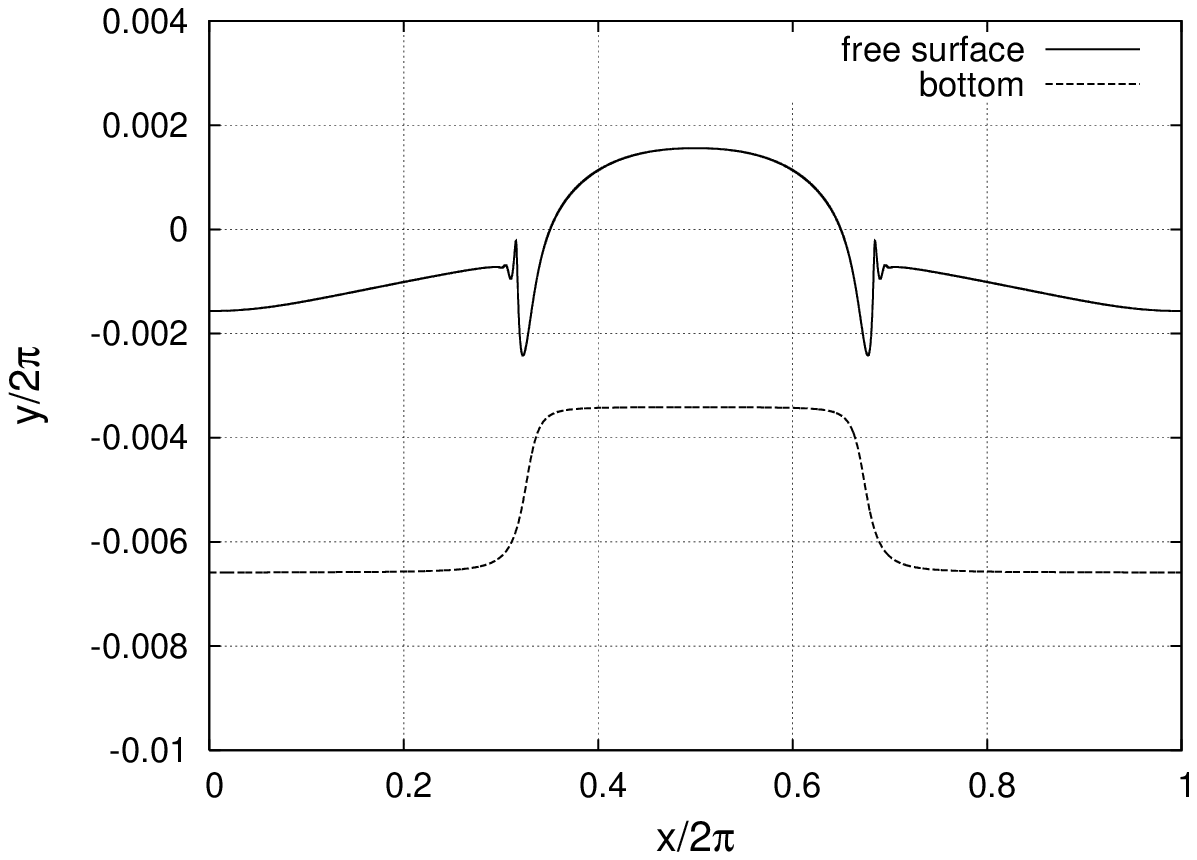,width=74mm}
  \epsfig{file=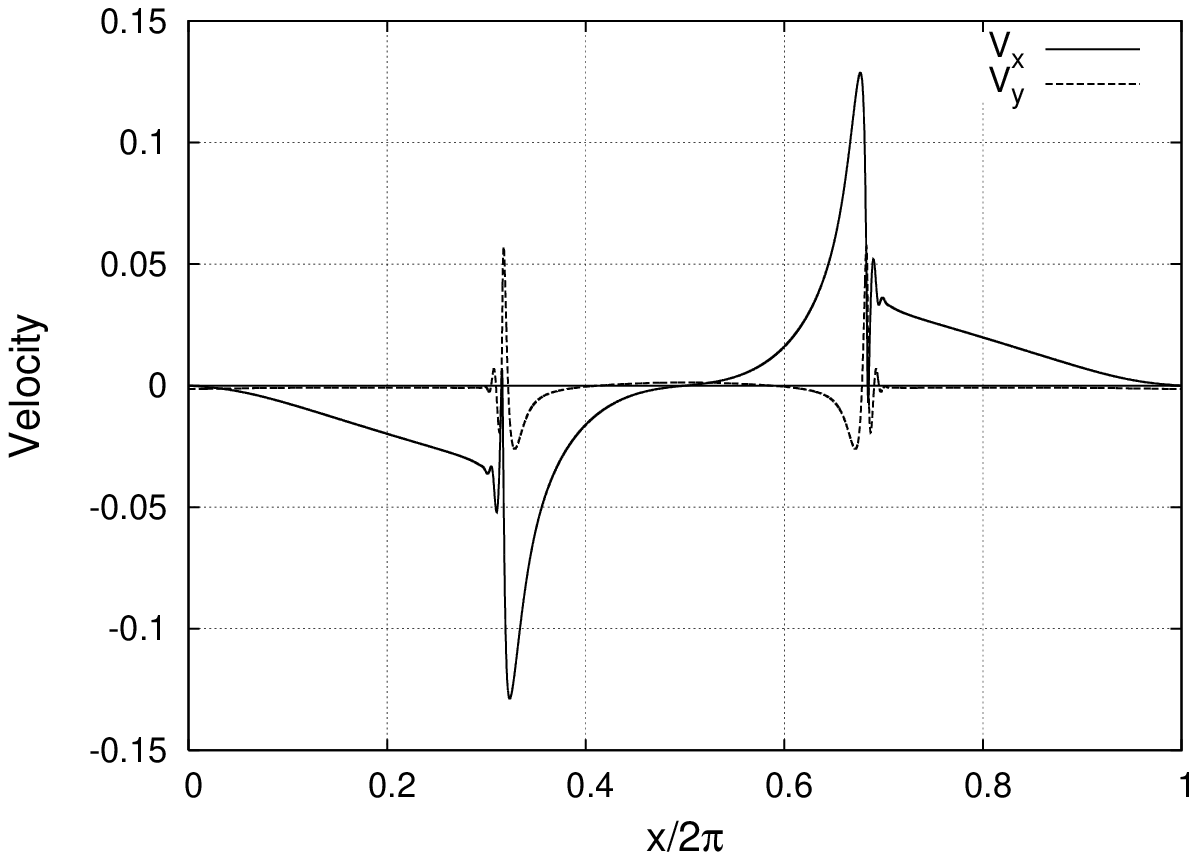,width=74mm}
\end{center}
\caption{\small (ii) $t=10$} 
\label{ii-t2}
\end{figure}
\begin{figure}
\begin{center}
  \epsfig{file=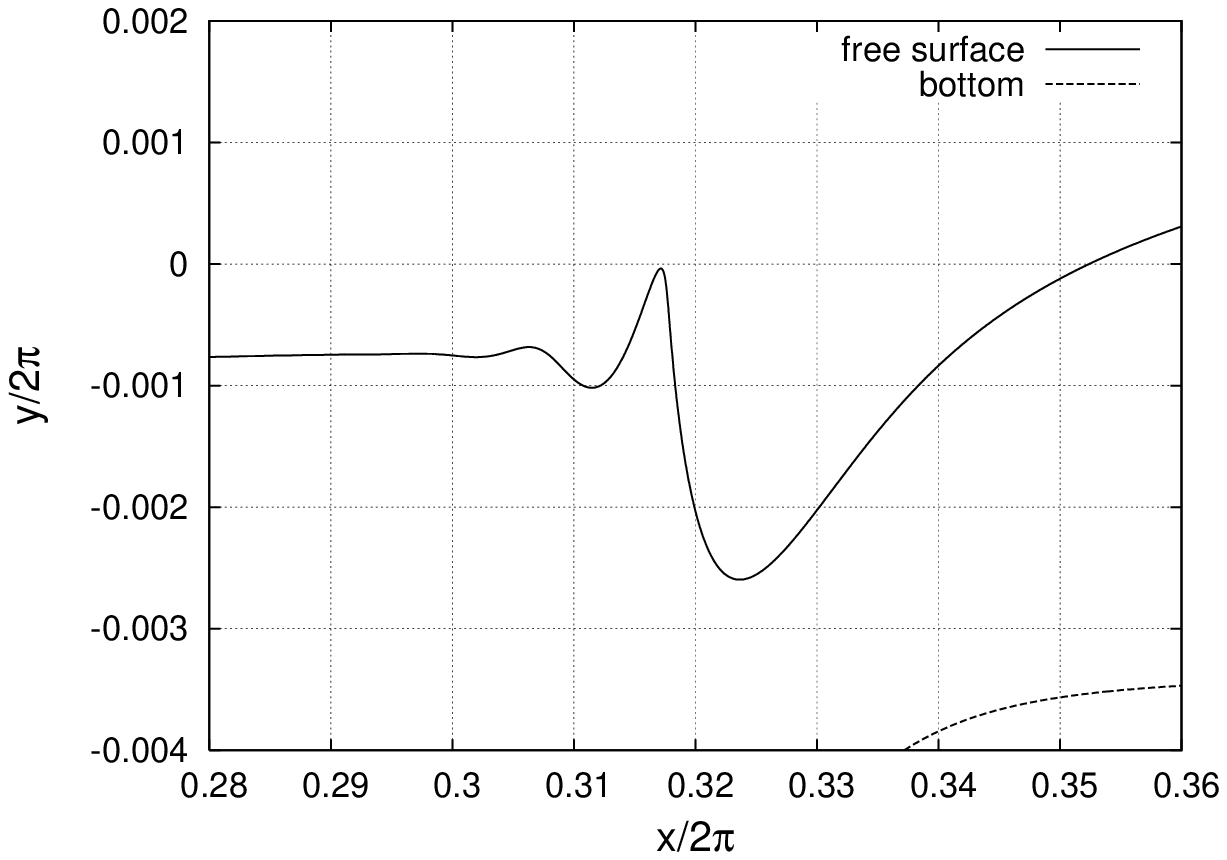,width=74mm}
  \epsfig{file=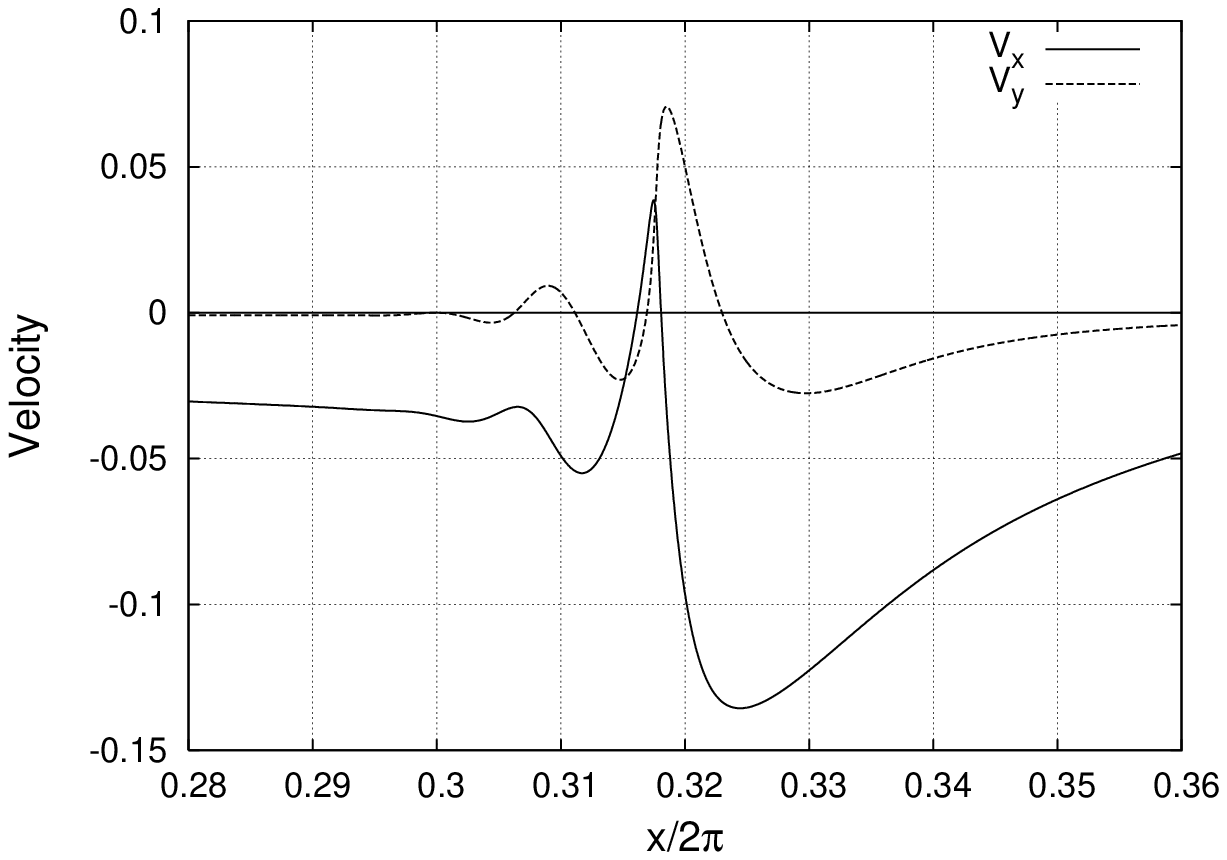,width=74mm}   
\end{center}
\caption{\small (ii) $t=10.117$} 
\label{ii-t3}
\end{figure}

In the second numerical experiment [referred to as (ii)],
the bottom profile was determined by
function $Z(\zeta,t)={\cal B}^{(ii)}(\zeta-i\alpha_0,t)$, where
\begin{equation} 
{\cal B}^{(ii)}(q,t)=q
-iA(\lambda t)^4e^{-\lambda t}
\ln\left(\frac{i\sin q+\sqrt{\epsilon+\cos^2q}}
{\sqrt{1+\epsilon}}\right),
\end{equation}
with $\alpha_0=\pi/100$, $\epsilon=0.01$, $A=0.1$, $\lambda=0.25$.
The initial values were
$\rho(\vartheta,0)=0$, $\chi(\vartheta,0)=0$, and $\alpha(0)=\alpha_0$, 
thus corresponding to the horizontal free surface $y=0$, 
the horizontal bottom $y=-\alpha_0$, and zero velocity field at $t=0$. 
At later times, some pieces of the bottom rose and some pieces went down.  
Results of this simulation are presented in Figs.\ref{ii-t1}-\ref{ii-t3}.
In this case, a singularity forms in the places where the fluid flows from
the shallow region to the deep region (see Fig.\ref{ii-t3}).

\subsection{Third numerical experiment}

\begin{figure}
\begin{center}
  \epsfig{file=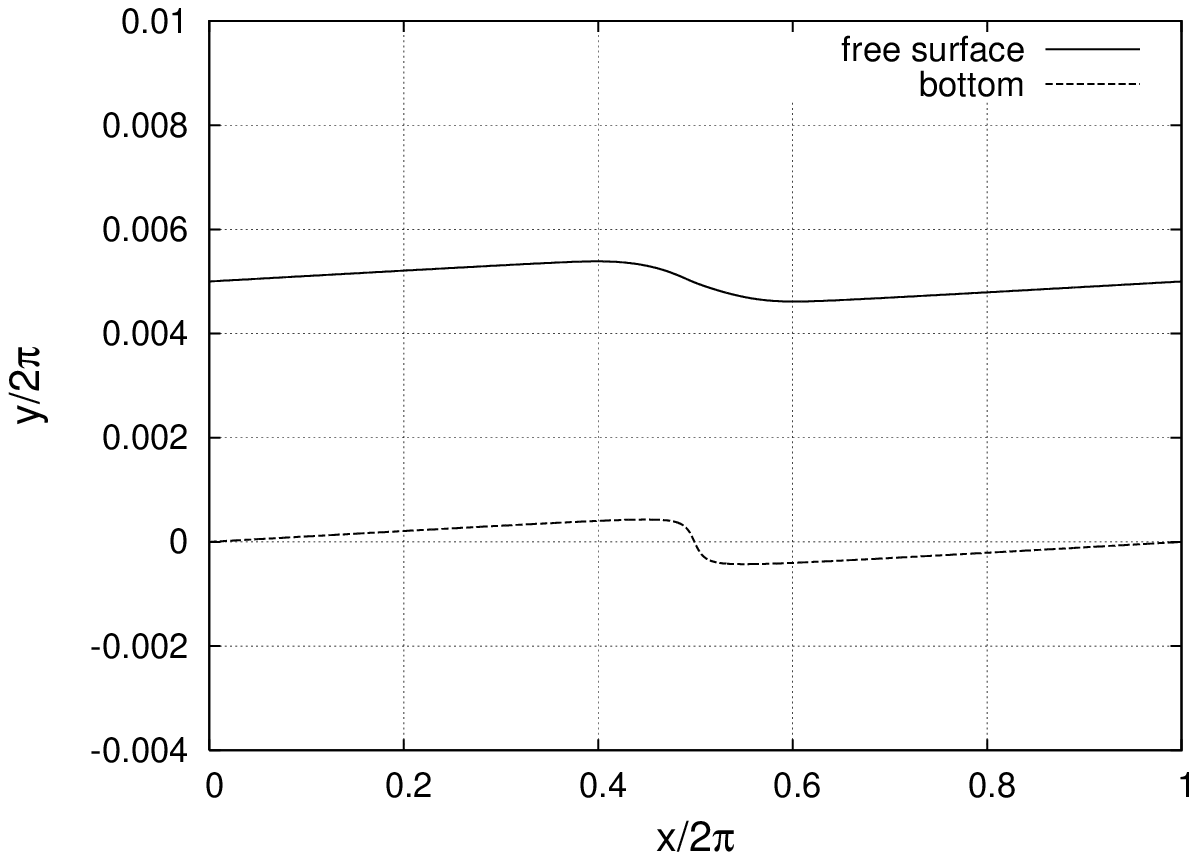,width=74mm}
  \epsfig{file=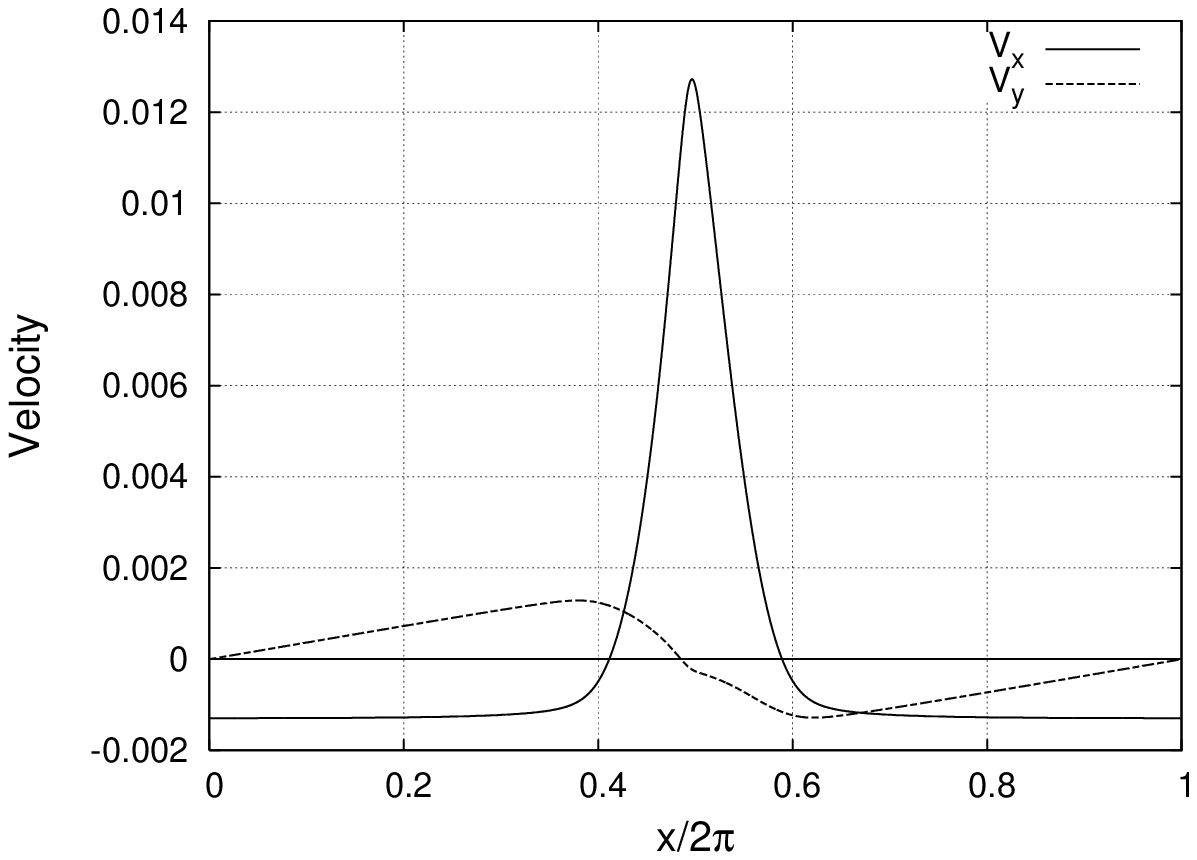,width=74mm}
\end{center}
\caption{\small (iii) $t=5$} 
\label{iii-t1}
\end{figure}
\begin{figure}
\begin{center}
  \epsfig{file=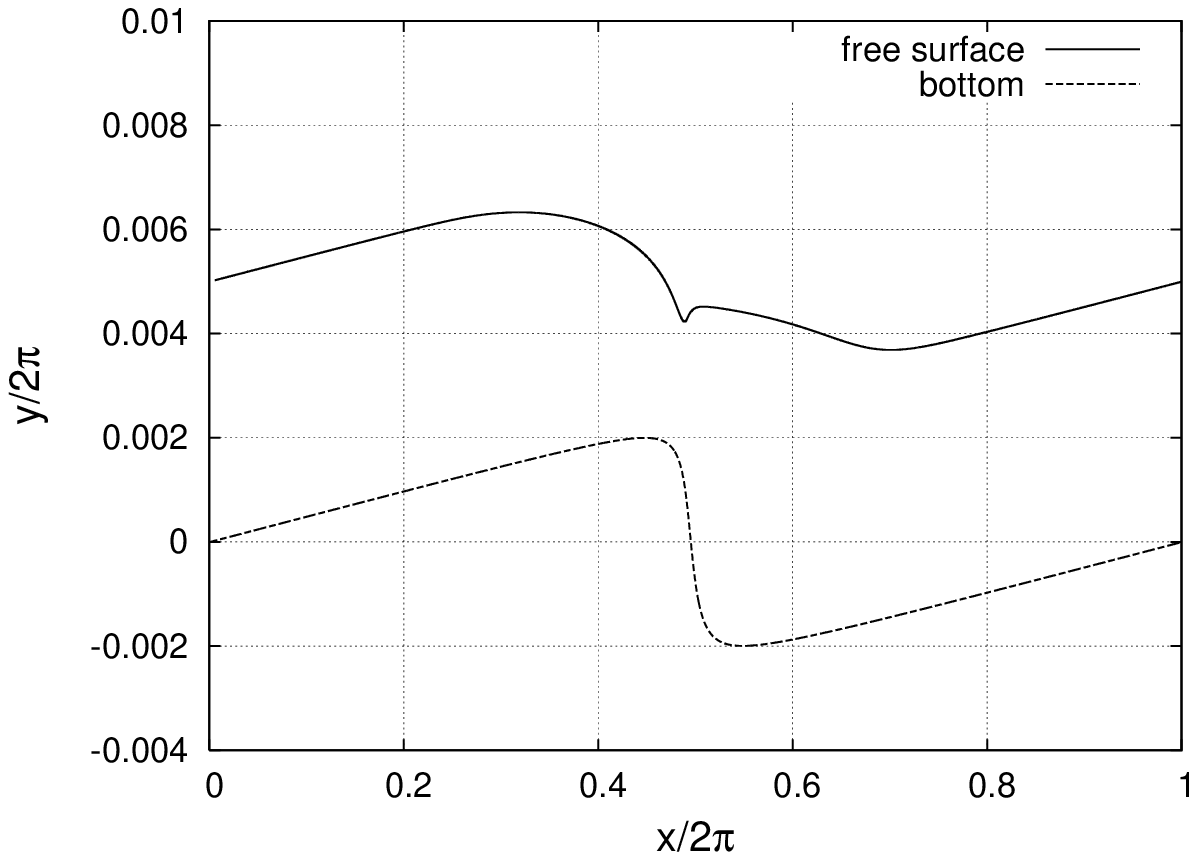,width=74mm}
  \epsfig{file=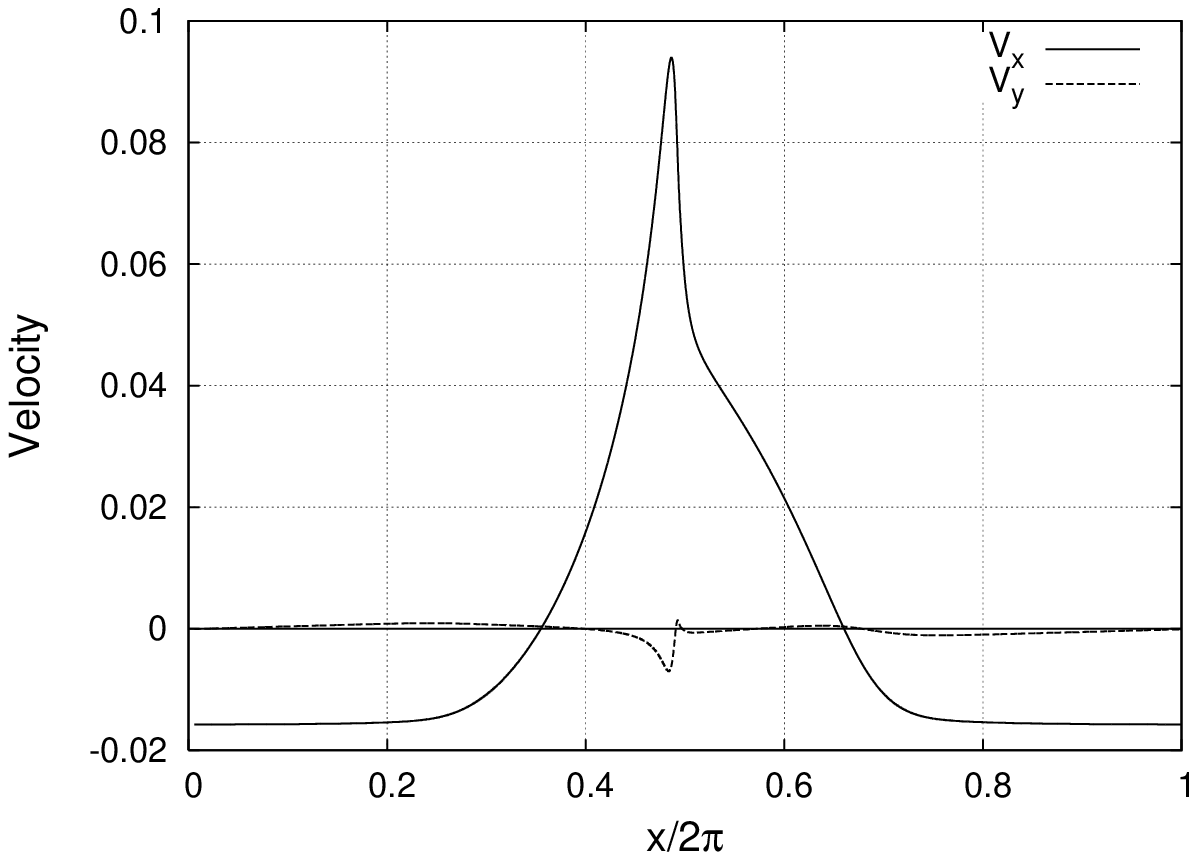,width=74mm}
\end{center}
\caption{\small (iii) $t=10$} 
\label{iii-t2}
\end{figure}
\begin{figure}
\begin{center}
  \epsfig{file=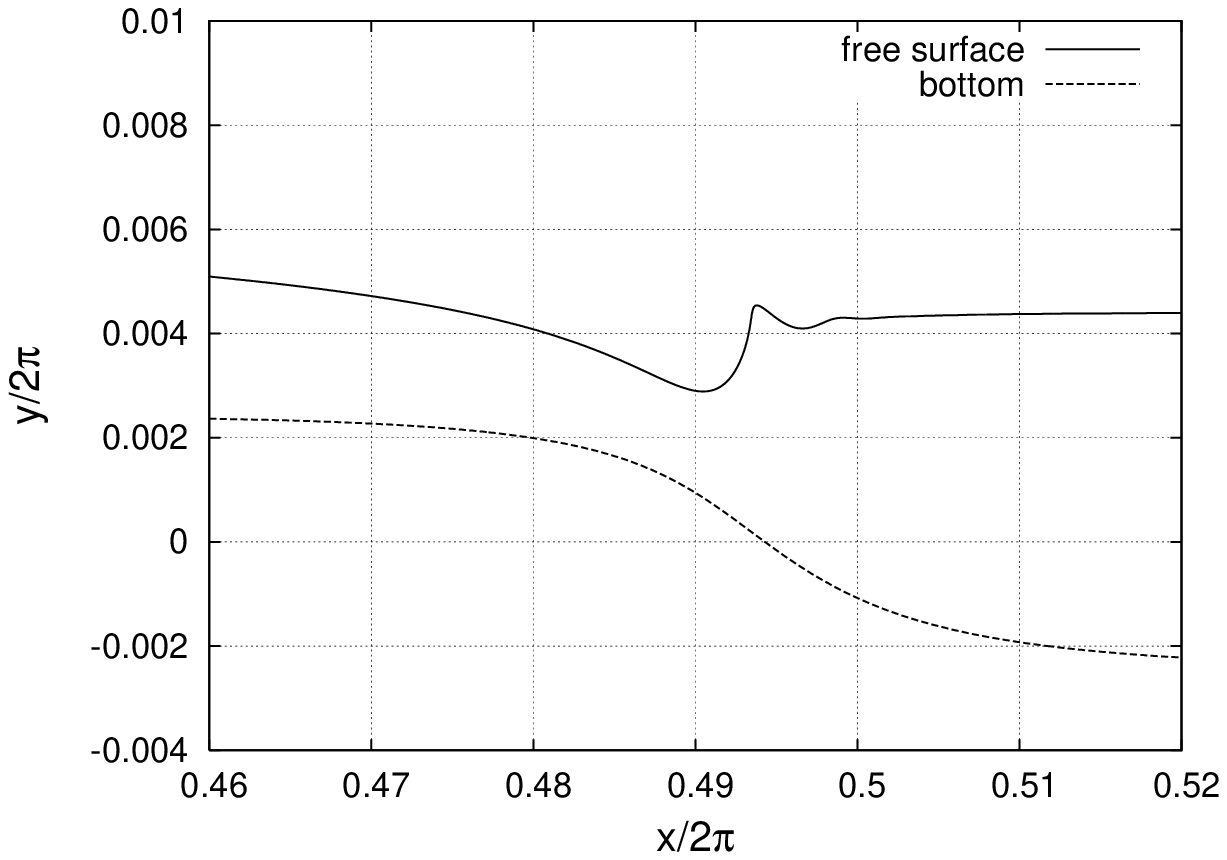,width=74mm}
  \epsfig{file=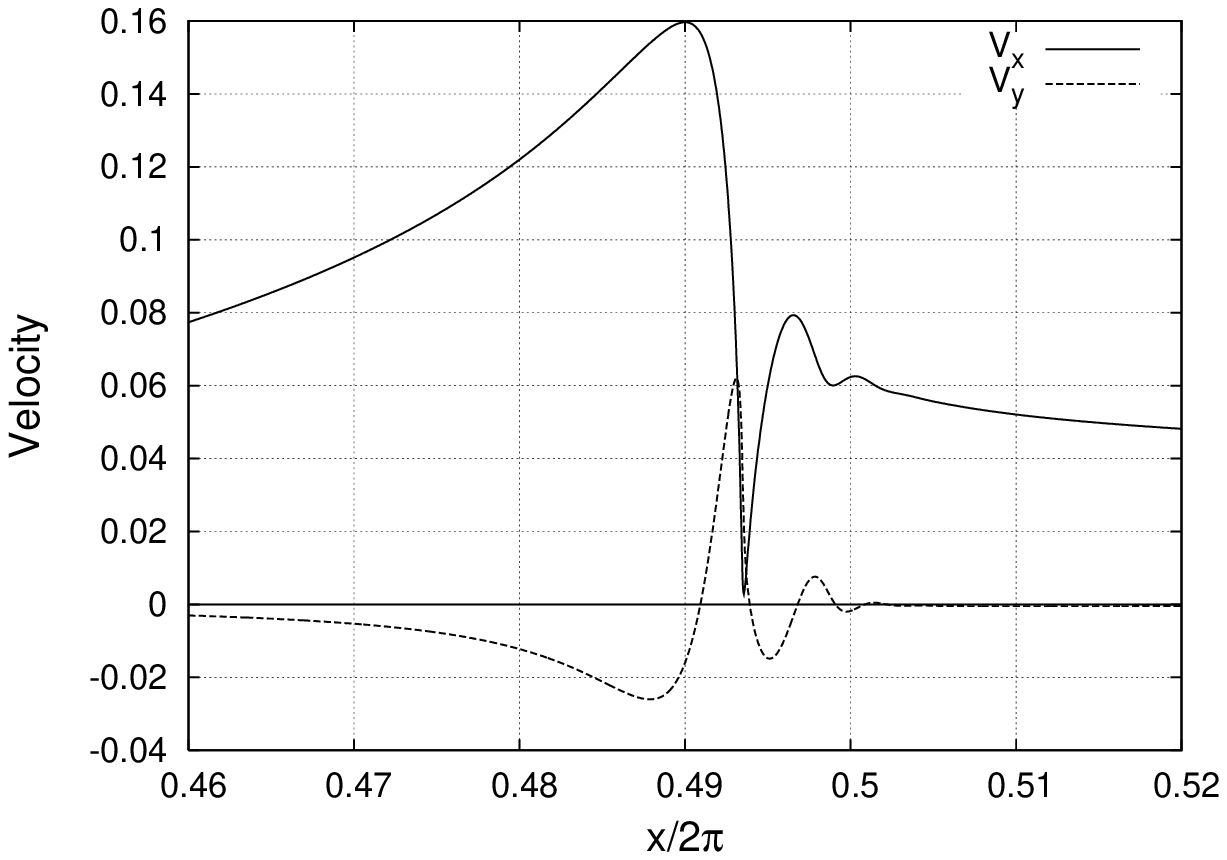,width=74mm}   
\end{center}
\caption{\small (iii) $t=11.40$} 
\label{iii-t3}
\end{figure}

In the third numerical experiment [referred to as (iii)], the following
function $Z(\zeta,t)$ was used:
\begin{equation} 
Z(\zeta,t)=\zeta+A(\lambda t)^4e^{-\lambda t}
\ln\left[1+C\exp(i\zeta)\right],
\end{equation}
with $A=\pi/1000$, $\lambda=0.25$, $C=0.95$. As previously, 
$\rho(\vartheta,0)=0$, $\chi(\vartheta,0)=0$, and $\alpha(0)=\alpha_0$.
The results of this simulation
are presented in Figs.\ref{iii-t1}-\ref{iii-t3}.

\section{Summary}

In this paper exact nonlinear equations of motion for potential 2D ideal
incompressible flows
with a free surface in a uniform gravitational field have been obtained, 
taking into account a time-dependent curved bottom with a profile prescribed 
by an analytical function. The mathematical theory of functions of a complex
variable and conformal mappings was extensively used in the derivation. 
Despite a complicated structure of the equations, an efficient numerical code 
has been developed for space-periodic solutions,
which gives very accurate results.


\end{document}